# Beating the fundamental rate-distance limit in a proof-of-principle quantum key distribution system

Shuang Wang,[1,2,3] De-Yong He,[1,2,3] Zhen-Qiang Yin,[1,2,3*] Feng-Yu Lu,[1,2,3] Chao-Han Cui[1,2,3], Wei Chen,[1,2,3†] Zheng Zhou,[1,2,3] Guang-Can Guo,[1,2,3] and Zheng-Fu Han[1,2,3]

1 *CAS Key Laboratory of Quantum Information, University of Science and Technology of China, Hefei 230026, China*

2 *CAS Center for Excellence in Quantum Information and Quantum Physics, University of Science and Technology of China, Hefei 230026, China*

3 *State Key Laboratory of Cryptology, P. O. Box 5159, Beijing 100878, China*

**With the help of quantum key distribution (QKD), two distant peers are able to share information-theoretically secure key bits. Increasing key rate is ultimately significant for the applications of QKD in lossy channel. However, it has proved that there is a fundamental rate-distance limit, named linear bound, which limits the performance of all existing repeaterless protocols and realizations. Surprisingly, a recently proposed protocol, called twin-field (TF) QKD can beat linear bound with no need of quantum repeaters. Here, we present the first implementation of TF-QKD protocol and demonstrate its advantage of beating linear bound at the channel distance of 300 km. In our experiment, a modified TF-QKD protocol which does not assume phase post-selection is considered, and thus higher key rate than the original one is expected. After well controlling the phase evolution of the twin fields travelling hundreds of kilometers of optical fibres, the implemented system achieves high-visibility single-photon interference, and allows stable and high-rate measurement-device-independent QKD. Our experimental demonstration and results confirm the feasibility of the TF-QKD protocol and its prominent superiority in long distance key distribution services.**

---

* yinzq@ustc.edu.cn

† weich@ustc.edu.cn

**Introduction**

Since the invention of the first QKD protocol [1] in 1984, great efforts have been devoted to improving its key rate in real lossy channel. To overcome the transmission loss of photons, which is the main obstacle to high key rate, the most powerful way may be introducing quantum repeaters [2-4], however, which are still far from real-life applications. In practice, many repeaterless QKD experiments [5-12] have been realized to increase the key rate and extend the channel distance. Nevertheless, the theorists have proposed that there are some fundamental limits [13,14] on the key rates of all these repeaterless QKD protocols and experiments. Denoting the transmission efficiency of the lossy channel as $\eta$, the key rate R for any point-to-point repeaterless QKD protocol will satisfy R $\leq -\log_2(1-\eta)$, i.e. R$\sim O(\eta)$ which is called linear bound [14]. Surprisingly, several months ago, a revolutionary work [15] pointed out that this bound may be overcome in a so-called twin-field (TF) QKD protocol. In TF-QKD protocol, both peers Alice and Bob prepare and send phase-coding optical fields (weak coherent states) to an untrusted third-party Charlie, who is in the middle of the channel and interfere the incoming fields. Then Alice and Bob can generate sifted key bits provided Charlie observes a single photon click after interference. One can imagine that if Charlie is honest, his counting rate of single photon click is only attenuated by the channel loss between Alice to Charlie or Bob to Charlie. Consequently, one may conjecture the R may be proportional to $\sqrt{\eta}$, thus R$\sim\sqrt{\eta}$ is expected. However, the full security of TF-QKD is not proved in Ref. [15], which leads to the questioning of its security and calculation of R [16]. Fortunately, subsequent theoretical works [17-22] remedied the security of TF-QKD with different methods and reconfirmed its advantage of beating linear bound. Besides, these theoretical works also showed that the security of TF-QKD does not rely on Charlie's measurement, thus is measurement-device-independent (MDI) [23].

Although the TF-QKD has attracted intensive theoretical studies, a successful demonstration of beating linear bound is still missing. This is partially because TF-QKD needs steady interference between two weak coherent states from distant peers. And, the original theory of TF-QKD predicts beating linear bound only occurs when channel distance is very long. Achieving steady interference between two sources so far away is very challenging, since the phase drift is more severe in longer channel distance. Besides, realizing and keeping two laser sources with high indistinguishability is also difficult.

Here, we firstly demonstrate a TF-QKD system, where beating linear bound is achieved at a channel distance over 300 km. Benefiting from TF-QKD without phase post-selection [20,21], our experiment exhibits key rate over linear bound at a channel distance of 300 km, which is much shorter than the minimum value to overcome the linear bound predicted by the original protocol. Besides, with no need of phase post-selection, our system enjoys a simple process of post-processing.

**Protocol**

A simplified version of TF-QKD is conducted in our experiment. Our protocol [20] consists of code mode and decoy mode. The former is used to generate sifted key bits, while the latter is used to collect some parameters to bound information leakage. The flow of our protocol can be summarized in four steps.

1. Code mode or decoy mode is randomly selected by Alice (Bob) in each trial.

2. In code mode, Alice and Bob prepare phase coding weak coherent states $| \pm \sqrt{\mu} >_A$ and $| \pm \sqrt{\mu} >_B$ respectively, then send them to Charlie who interference the incoming weak coherent states and measures the phase shift between them. If Charlie successes to obtain the phase shift, Alice and Bob will retain this key bit. According to Charlie's measurement result, Bob may decide to flip his key bit or not.

3. In decoy mode, which is quite similar with decoy-state method [24-26] used in MDI-QKD protocol. Alice and Bob prepare and send phase randomized weak coherent states with four different intensities ($\mu, \nu_1, \nu_2, \nu_3$). Charlie is not aware of code or decoy modes. He still performs measurement on phase shift and announces his measurement result to Alice and Bob.

4. After repeating steps 1-3 for many times and some public communications, Alice and Bob can accumulate sufficient sifted key bits from code modes and estimate the yields for decoy states $Q_d^{xy}$ ($x, y = \mu, \nu_1, \nu_2, \nu_3$). For instance, $Q_d^{\mu\nu_1}$ is the probability that Charlie announces a successful measurement on phase shift when Alice and Bob actually prepare weak coherent pulses with mean photon number of $\mu$ and $\nu_1$ respectively in the decoy mode. From $Q_d^{xy}$, information leakage can be bounded, and then secret key bits may be generated.

A notable advantage of our protocol is phase randomization and post-selection in code mode are both removed. Thus, experimental system is simplified, and higher key rate is expected.

**Implementation system of TF-QKD**

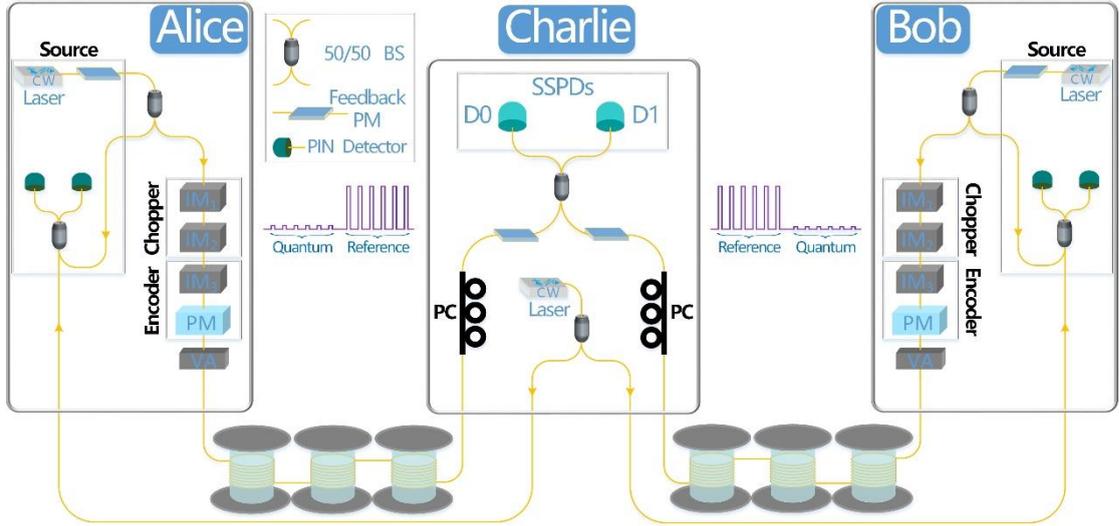

**Figure 1. Experimental setup to implement the twin-field quantum key distribution**. CW Laser: continuous- wave laser; IM: intensity modulator; PM: phase modulator; VA: variable attenuator; 50/50 BS: beam-splitter with a 50/50 splitting ratio; PC: polarization controller; SSPDs: superconducting single photon detectors. Alice and Bob have the same setup, which mainly consists of the source, chopper and encoder modules. The source generates the laser with a locked frequency and phase with the laser from Charlie. The chopper is used to first modulate the CW laser into a pulse train (IM$_1$), and then chop it into time-multiplexed bright reference and weak quantum parts (IM$_2$). The encoder adds code or decoy information into the quantum part (IM$_3$ and PM).

The experimental setup performing the TF-QKD protocol is summarized in Fig. 1. As a MDI scenario, two senders Alice and Bob have symmetric positions in relation to the measurement node Charlie. Alice and Bob have the same experimental setup, which mainly consists of three modules denoted as source, chopper and encoder, respectively. Both Alice's and Bob's **sources** are phase locked with the laser from Charlie, to generate the twin fields with a central wavelength of 1550.12 nm.

**The chopper** is composed of two intensity modulators (IM), IM$_1$ first modulates the locked continuous-wave (CW) laser into a pulse train with a 130 ps temporal width at a repetition rate of 1 GHz, IM$_2$ then chops the pulse train into the time-multiplexed reference part and quantum part, in which the reference part is bright and unmodulated by the encoder in order to measure the phase shift of the channels, and the quantum part carries the information of the keys, the duration time of either part is 50 μs.

**The encoder** only applies to the quantum part, and is composed of IM$_3$ and one phase modulator (PM), IM$_3$ is used to create four intensity levels required by the protocol, PM is used to modulate specific or random phase on each pulse of the quantum part. The encoder is randomly operated in the code mode or decoy mode. In the code mode, IM$_3$ creates the signal state with μ photons per pulse, and PM modulates the phase $\{0, \pi\}$ according to the random key bit $\{0, 1\}$. In the decoy mode, $IM_3$ randomly creates four decoy states with μ, $\nu_1$, $\nu_2$, and $\nu_3$ photons per pulse, and PM randomizes the phase of each pulse belonged to the quantum part with amplitude resolution of 10 bits [27,28].

**Charlie** is in the middle to take a single-photon interference measurement. The two fields sent by Alice and Bob interference on the 50/50 beam-splitter (BS). In order to achieve a good interference visibility, a polarization controller (PC) and a feedback PM are added before each input of the BS. The PC is used to set correct polarization, and the feedback PM is used to compensate the fast phase drift in fibre channels. The interference results are detected by two superconducting single photon detectors (SSPDs). When both Alice's and Bob's encoders are in the code mode, the detector D0 would click if the phase difference modulated by Alice and Bob is 0, and the detector D1 would click if the phase difference modulated is $\pi$. These two SSPDs are made by Scontel Inc., their detection efficiency is more than 60%, and total dark count rate is less than 200 Hz.

## Experimental results

There are two main technical challenges when TF-QKD is implemented. One is making the fields generated by Alice and Bob 'twin', which means the frequency difference between Alice' s and Bob's lasers is 0, and the phase difference between these two lasers is stable. The other is compensation of the fast phase drift over long fibers.

### Results of the source part

The optical phase-locked loop (OPLL) architecture [29,30] is employed in the source part to generate the twin fields between Alice and Bob. The CW laser transmitted from Charlie takes the role of a master laser, while the local laser belonged to Alice (Bob) takes the role of a slave laser. The master laser and part output of the slave laser are combined in a 50/50 BS to produce a phase error signal, then detected by two PIN detectors that is connected to the feedback loop. The error signal from PIN detectors passes through a loop filter, generates a bias current to tune the frequency of the slave laser. Once the OPLL is locked, the phase difference between the master and slave lasers is stable, and these two lasers have the same frequency.

The master laser (X15 model) is an ultralow noise fibre laser made by NKT Photonics Inc. During all measurements in the experiment, the master laser is free running, in which its linewidth is less than 0.1 kHz and wavelength stability over case temperature is approximately 0.1 pm/℃. The slave laser is an external cavity diode laser with a standard 14-pin butterfly package, whose driving circuit is homemade. The external cavity diode laser (PLANEX™, RIO Inc.) has a linewidth of less than 2 kHz, and wavelength sensitivity to temperature and bias current of approximately 12.5 pm/℃ and 0.2 pm/mA, respectively. The resolution of the laser temperature controller is 0.001 ℃. In the experiment, the central wavelength of the master laser is set at 1550.12nm, the wavelength of the slave laser is slowly tuned by changing its temperature. Once the beat note frequency of these two laser is close to 10 MHz (0.08 pm), the homodyne OPLL will be locked, and finally achieve a zero offset frequency between the master and slave lasers.

The lock performance is tested by observing the beat note signal and the interference visibility, respectively. The beat note signal is measured with a PIN detector and recorded with an oscilloscope. The result of the quadrature interference is displayed in Fig. 2a, and exhibits stable interference result. The interference visibility is measured with two SSPDs, one connects the constructive interference output, and the other connects the destructive interference output. The counts of these two SSPDs are shown in Fig. 2c, each point is the total counts in 50 μs, and the curve is smoothened by averaging every 20 adjacent points. The distributions of the constructive count and destructive count are shown in the insets of Fig. 2c, respectively. The mean value of the constructive count is 251.97, and of the destructive count is 2.26, so the interference visibility is approximately 98.22%.

Furthermore, a feedback PM is added in the source part to reduce residual phase noise. The error signal from PIN detectors is amplified and connected to the PM as negative feedback. The corresponding result of the quadrature interference is displayed in Fig. 2b, the improvement is significant compared with Fig. 2a. With almost the same mean value of 32 mV, the standard deviation is changed from 3.18 mV to 0.92 mV. The corresponding counts of two SSPDs are shown in Fig. 2d. Compared with Fig. 2c, the improvement mainly comes from the destructive count, whose mean value is changed from 2.26 to 0.27, the corresponding interference visibility becomes approximately 99.78%.

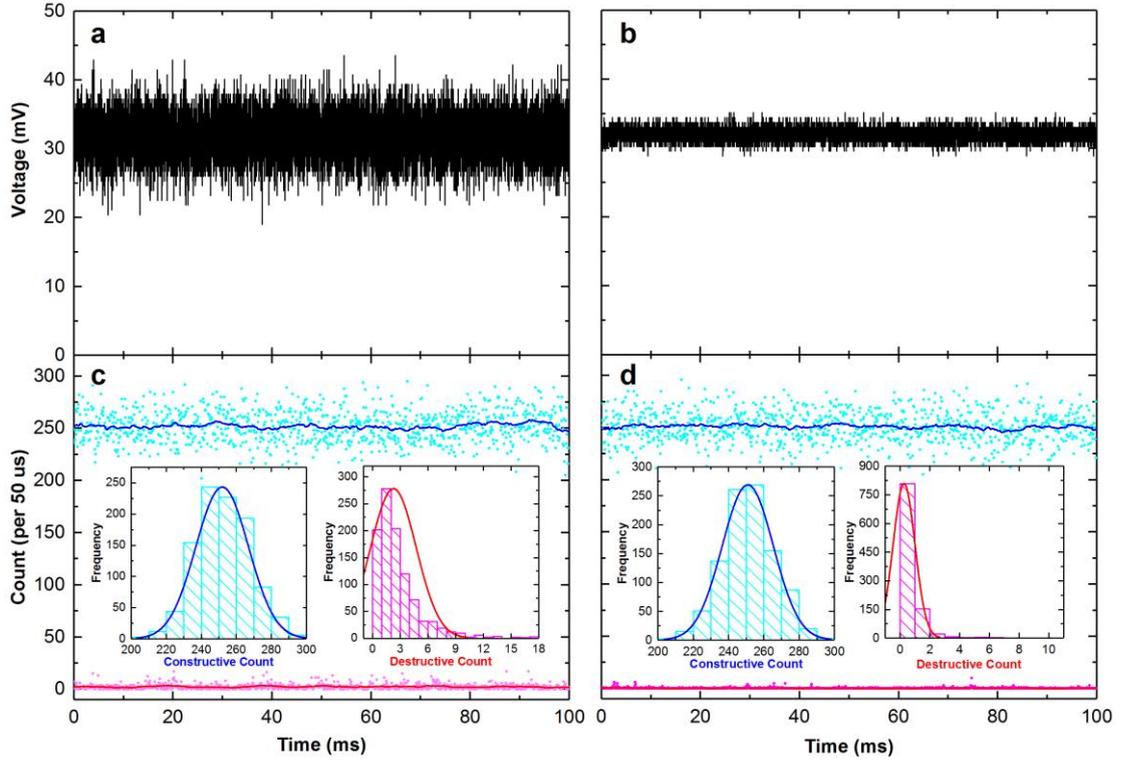

**Figure 2. Characteristics of the source part.** a. The quadrature interference results recorded by an oscilloscope. b. The quadrature interference results with a feedback PM. c. The in-phase interference results detected by two SSPDs. d. The in-phase interference results with a feedback PM. The cyan points correspond to the counts of the constructive interference output, the magenta points refer to the counts of the destructive interference output. The curves are smoothened by averaging every 20 adjacent points. The insets are the distributions of the constructive counts and destructive counts.

## Results of compensation of the fast phase drift

After making the fields generated by Alice and Bob 'twin', the phase drift is mainly due to the fluctuations in fiber channels. Depending on the surrounding environment (the temperature and vibration), the phase drift accumulates with the length of fibre channels. To compensate this fast phase drift over much longer fibres, a feedback PM is inserted in each arm of the BS at Charlie's site. The active feedback loop that acts on the feedback PMs is based on the interference outputs of Alice's and Bob's bright pulses belonged to the corresponding reference parts, which are unmodulated and time-multiplexed with quantum parts. The interference outputs are detected by two SSPDs (D0 and D1). The goal of the active feedback loop is to maximize (minimize) the counts of the detector D0 (D1) through fast change the voltage of the feedback PMs. Based on the counts of D0 and D1 belonged to the reference part, the active feedback loop estimates the corresponding compensation voltage, and immediately loads it on the feedback PMs.

The active feedback is realized with a field programmable gate array (FPGA), operating at a 40 MHz clock-rate that is synchronized with the QKD control system. The feedback PM with a insertion loss of 2.2 dB has a high impedance input for a bandwidth of approximately 200 MHz. Considering the actuating time from FPGA to PM ($\sim$ 0.2 µs) and the transition time from reference part to quantum part, a time window of 48 µs is set to select the counts during each 50 µs.

The performance of the compensation of the fast phase drift is tested without and with the active feedback loop. During the test, both Alice's and Bob's sources have been locked with the laser from Charlie, the chopper works, and only the counts belonged to the reference part are recorded. At a total distance of 300 km optical fibre, the pulses in the reference part is attenuated to approximately 2.9 photons per pulse. The phase drift is measured when the driving voltage over the feedback PMs is disconnected, and the counts of the detectors D0 and D1 are displayed in Fig. 3a. The phase drift mainly depends on the ambient vibration, the drift rate in this 100 ms is less than $\pi$ rad/ms. The results of active compensating the phase drift is shown in Fig. 3b. After connecting the PMs' driving voltage, the counts of D0 and D1 stay relatively stable, D0 corresponds to the constructive interference and D1 corresponds to the destructive interference. The mean value of the count of D0 is 186.66, and of the count of D1 is 2.64, so the interference visibility is approximately 97.21%.

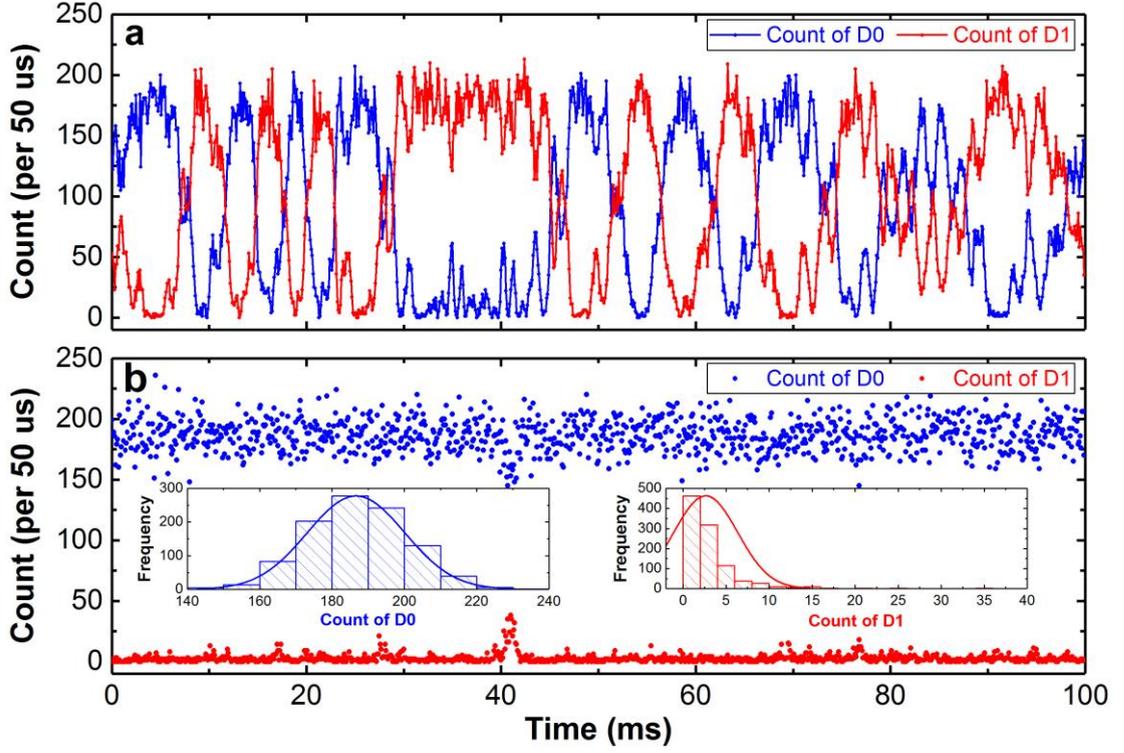

**Figure 3. Characteristics of the compensation of the fast phase drift at a total distance of 300 km optical fibre.** a. The phase drift by recording the counts of D0 and D1 when the driving voltage over the feedback PMs is disconnected. b. The results of active compensating the phase drift by recording the counts of D0 and D1 after connecting the PMs' driving voltage. The insets are the distributions of the counts of two SSPDs. Here only the counts belonged to the reference parts are recorded.

**Performance of the TF-QKD system**

After the main technical challenges are tackled, the TF-QKD experiment is performed over three total distances of 100 km, 200 km, and 300 km optical fibre. The optical fibres used in the experiment are standard single mode fibres (ITU-G. 652D) with a loss coefficient of approximately 0.18 dB/km. The overall loss of Charlie's deices is approximately 5.16 dB. From the yields $Q_{\mu\mu}$, $Q_d^{\mu\mu}$, $Q_d^{v_1v_1}$, $Q_d^{v_2v_2}$, $Q_d^{v_3v_3}$, $Q_d^{\mu v_3}$, $Q_d^{v_1v_3}$, $Q_d^{v_2v_3}$, $Q_d^{v_3\mu}$, $Q_d^{v_3v_1}$, $Q_d^{v_3v_2}$ and the error rate $e_b$, the upper bound of information leakage $I_{AE}^u$ could be estimated numerically (see Appendix). Then the secret key rate is obtained by $R = Q_{\mu\mu}(1 - fh_2(e_\mu) - I_{AE}^u)$, where the efficiency of error correction $f = 1.15$ and $h_2$ is the binary Von Neumann entropy. The results are listed in the following table (Tab. I). For comparison, the linear bound of secret key rate $R_{LB}$ is also presented. We can clearly see that our secret key rate overwhelms the linear bound at a channel distance of 300 km.

**Tab.I Experimental results**

| Distance (km) | Total loss (dB) | Visibility | $Q_{\mu\mu}$ | $e_\mu$ | $R$ | $R_{LB}$ |
|---|---|---|---|---|---|---|
| 100 | 23.06 | 98.64% | $2.02 \times 10^{-3}$ | 1.86% | $8.87 \times 10^{-4}$ | $7.15 \times 10^{-3}$ |
| 200 | 40.66 | 98.05% | $1.94 \times 10^{-4}$ | 2.42% | $8.01 \times 10^{-5}$ | $1.24 \times 10^{-4}$ |
| 300 | 58.46 | 96.79% | $2.11 \times 10^{-5}$ | 3.59% | $6.46 \times 10^{-6}$ | $2.06 \times 10^{-6}$ |

The key rate of our implementation, the simulation results and the linear bound are summarized in Fig.4. The key rates of simulation and linear bound are plotted against the total distance of fibre channels (from Alice to Bob) with a loss efficiency of 0.18 dB/km. (Details of the simulation see Appendix). Experimental results (Open dots) at the distances of 100 km, 200 km, and 300 km are moved to the positions with equivalent attenuations. For instance, the measured loss of 300 km fibre in the experiment is approximately 53.3 dB, the corresponding experimental dot is moved to the distance of 296 km in Fig.4. In general, the experimental points fit the simulation results quite well. In the same environment, the phase drift of the channels mainly depends on the length of optical fibres. Employing the same active feedback module, the phase drift could be compensated very well at relatively short distances. The visibility of the reference parts is 98.64% at the distance of 100 km, and 98.05% at the distance of 200 km. Due to relatively high visibility of reference parts and low QBER of quantum parts, the key rates at the distances of 100 km and 200 km are a little higher than the simulation results.

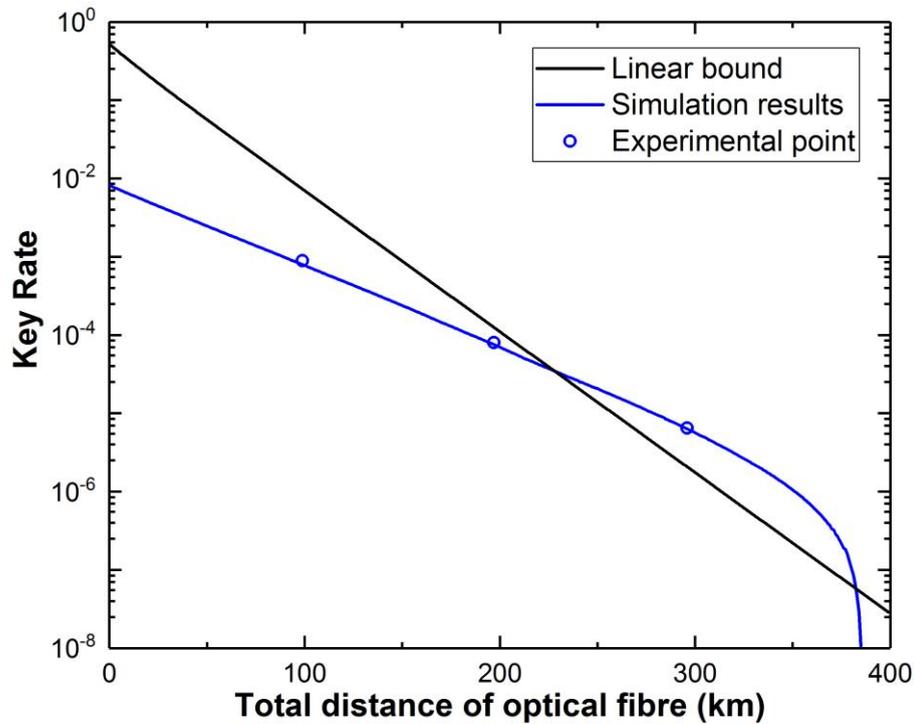

**Figure 4. Key rate of the TF-QKD system.** The rates are plotted against the total distance of optical fibre (from Alice to Bob) with a loss efficiency of 0.18 dB/km. The open dots refer to experimental points at distances of 100 km, 200 km, and 300 km, respectively.

## Discussion

At a total distance of 300 km optical fibre, the stability of the TF-QKD system is shown with the interference visibility (blue open dots) and QBER (red open dots) in Fig.5. Over 1000 seconds, the mean value of the interference visibility of Alice's and Bob's reference parts is 96.86%, and the mean value of the QBER of corresponding quantum parts is 3.56%. Since the phase drift mainly depends on the ambient vibration, there are some abrupt vibrations that cannot be compensated very well, and cause some relatively low visibility and high QBER, such as the dots at 182nd second, the interference visibility and QBER are 94.54% and 4.58%, respectively. Still, the performance of the system is relatively stable, the standard deviation of the interference visibility and QBER are 0.28% and 0.23%, respectively.

Compared with the interference visibilities at the distances of 100 km and 200 km (above 98%), the relatively low visibility at the distance of 300 km shows the limitation of the compensation of the phase drift. If we want to reduce the channel optical error, the interference visibility of the reference parts should be improved. The duration time of both reference and quantum parts need to be shortened, and the intensity of the reference part need to be increased. That's a key step towards longer distance, in addition to choosing a large misalignment error tolerant protocol [18] and SSPDs with ultra-low dark count rate.

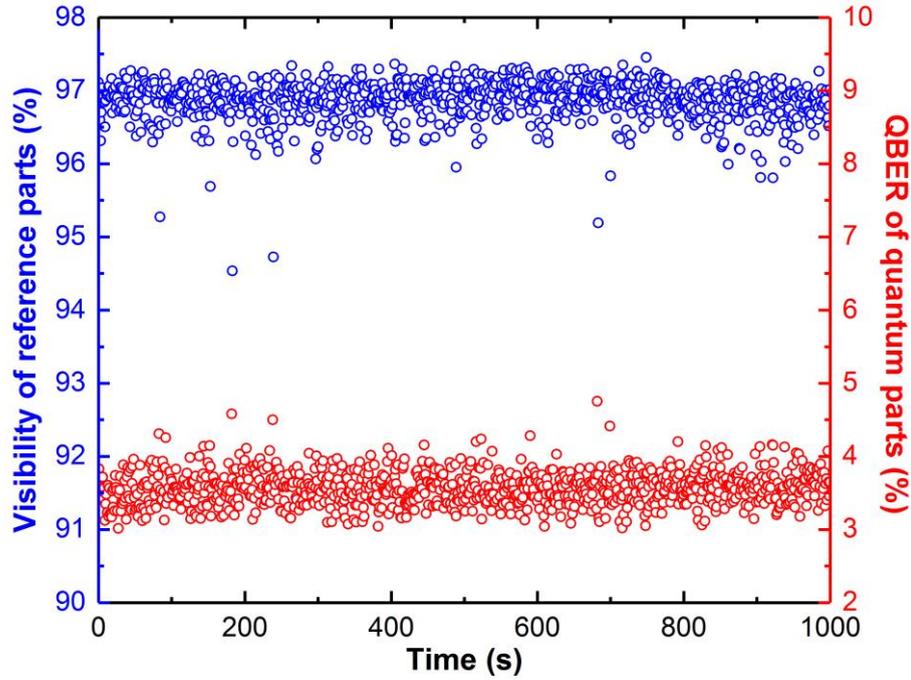

**Figure 5. Stability of the TF-QKD system over a total distance of 300 km.** The blue open dots represent the interference visibility of Alice's and Bob's reference parts, and the red open dots are QBER of corresponding quantum parts. Each dot corresponds to the data acquired in one second.

To summarize, we have successfully demonstrated an implementation of the TF-QKD protocol without phase post-selection. The implemented system can well control the phase evolution of the twin fields travelling hundreds of kilometers of optical fibre channels to achieve a single-photon interference with high visibility. Moreover, at a total distance of 300 km standard single mode fibre, our system overcomes the fundamental rate-distance limit of QKD. To our knowledge, this demonstration is the first QKD experiment that does beat the well-known linear bound of secret key rate. The system runs a modified version of the original TF-QKD. Its code mode no longer requires phase randomization and post-selection, thus secret key rate is further improved compared with the original TF-QKD. Our achievement demonstrates that TF-QKD protocol is feasible in practice, and will be a very promise solution for high-rate QKD over long distance in the future. We believe that there will be a notable improvement on the performance of QKD products with the help of TF-QKD protocol. However, there are still several points that must be addressed in future study. First one is the finite-key

## Acknowledgements

The first two authors, Shuang Wang and De-Yong He, contributed equally to this work. This work has been supported by the National Key Research and Development Program of China (2016YFA0302600), the National Natural Science

# Appendix

**The calculation of key rate** The key rate is obtained by $R = Q_{\mu\mu}(1 - f h_2(e_b) - I_{AE}^u)$. The essential thing is to estimate the upper bound of Eve's information on key bits $I_{AE}^u$. According to Eq.(2) of Ref. [20], $I_{AE}^u$ can be calculated by an optimization problem, which is

$$I_{AE}^u = \max_x h\left(\frac{x_{00}}{Q_{\mu\mu}}, \frac{x_{10}}{Q_{\mu\mu}}\right) + h\left(\frac{x_{11}}{Q_{\mu\mu}}, \frac{x_{01}}{Q_{\mu\mu}}\right).$$

The constraints for the non-negative real variables $x_{00}$, $x_{10}$, $x_{11}$ and $x_{01}$ are functions of $Y_{n,m}$, which is defined as yield when Alice and Bob prepare $n$-photon and $m$-photon respectively in decoy mode. In the case of finite decoy states, the upper bound and lower bound of $Y_{n,m}$, i.e. $Y_{n,m}^u$ and $Y_{n,m}^l$, can be used to bound $x_{00}$, $x_{10}$, $x_{11}$ and $x_{01}$. Then $I_{AE}^u$ is obtained. Concretely, linear programming is used to search $Y_{0,0}^{u(l)}$, $Y_{0,1}^{u(l)}$, $Y_{1,0}^{u(l)}$, $Y_{1,1}^{u(l)}$, $Y_{0,2}^{u(l)}$, and $Y_{2,0}^{u(l)}$ satisfying the experimental observed yields $Q_d^{\mu\mu}$, $Q_d^{v_1 v_1}$, $Q_d^{v_2 v_2}$, $Q_d^{v_3 v_3}$, $Q_d^{\mu v_3}$, $Q_d^{v_1 v_3}$, $Q_d^{v_2 v_3}$, $Q_d^{v_3 \mu}$, $Q_d^{v_3 v_1}$, and $Q_d^{v_3 v_2}$, since $Q_d^{xy} = \sum_{n,m=0}^{+\infty} P_n^x P_m^y Y_{n,m}$, where $P_n^x = e^{-x} x^n / n!$. Similarly, the lower bound of $P_0^\mu P_0^\mu Y_{0,0} + P_0^\mu P_1^\mu Y_{0,1} + P_1^\mu P_0^\mu Y_{1,0} + P_2^\mu P_0^\mu Y_{2,0} + P_0^\mu P_2^\mu Y_{2,0} + P_1^\mu P_1^\mu Y_{1,1}$, is obtained by linear programming. With these bounds, the constraints of $x_{00}$, $x_{10}$, $x_{11}$ and $x_{01}$ are established by Eqs.A.22-A.25 of Ref. [20]. Finally, $I_{AE}^u$ and key rate $R$ are found.

**The detailed experimental data**

We list the detailed experimental data in the following two tables.

| L(km) | Att(dB) | μ | $v_1$ | $v_2$ | $v_3$ | Q | e(%) |
|---|---|---|---|---|---|---|---|
| 100 | 17.9 | 0.026 | 0.005 | 0.002 | $8 \times 10^{-5}$ | $2.02 \times 10^{-3}$ | 1.86 |
| 200 | 35.5 | 0.019 | 0.005 | 0.002 | $6 \times 10^{-5}$ | $1.94 \times 10^{-4}$ | 2.42 |
| 300 | 53.3 | 0.016 | 0.005 | 0.002 | $5 \times 10^{-5}$ | $2.11 \times 10^{-5}$ | 3.59 |

Table I. L: the distance between Alice and Bob, Att: attenuation of channel between Alice and Bob, μ: the intensity of signal state, $v_1(v_2, v_3)$: the intensities of decoy states, Q: the yield of code mode, e: the error rate of raw key bit.

| L(km) | $Q_d^{\mu\mu}$ | $Q_d^{v_1 v_1}$ | $Q_d^{v_2 v_2}$ | $Q_d^{v_3 v_3}$ | $Q_d^{\mu v_3}$ ($Q_d^{v_3 \mu}$) | $Q_d^{v_1 v_3}$ ($Q_d^{v_3 v_1}$) | $Q_d^{v_2 v_3}$ ($Q_d^{v_3 v_2}$) |
|---|---|---|---|---|---|---|---|
| 100 | $2.02 \times 10^{-3}$ | $3.88 \times 10^{-4}$ | $1.56 \times 10^{-4}$ | $6.40 \times 10^{-6}$ | $1.01 \times 10^{-3}$ | $1.97 \times 10^{-4}$ | $8.10 \times 10^{-5}$ |
| 200 | $1.94 \times 10^{-4}$ | $5.12 \times 10^{-5}$ | $2.06 \times 10^{-5}$ | $8.02 \times 10^{-7}$ | $9.73 \times 10^{-5}$ | $2.60 \times 10^{-5}$ | $1.07 \times 10^{-5}$ |
| 300 | $2.11 \times 10^{-5}$ | $6.74 \times 10^{-6}$ | $2.81 \times 10^{-6}$ | $2.55 \times 10^{-7}$ | $1.07 \times 10^{-5}$ | $3.50 \times 10^{-6}$ | $1.53 \times 10^{-6}$ |

Table II. The yields of decoy mode.

**The simulation of key rate** In the simulation, we assume the transmittance of channel is $\eta = 10^{-0.018l} \eta_D$, in which $l$ is the channel distance (km) and $\eta_D = 0.305$ is the overall efficiency of measurement device. The dark count rate of each channel of SPD is $d = 10^{-7}$ per pulse. The optical misalignment is set to be 0.03. In the decoy mode, $v_1 = 0.005$, $v_2 = 0.002$ and $v_3 = 10^{-2.5}\mu$, while $\mu$ is optimized to maximize the key rate at each distance. All experimental observed yields and error rate can be simulated with the formulae given in the appendix B of Ref. [20].